\documentclass[twocolumn,showpacs,preprintnumbers,amssymb,aps, prl]{revtex4}
\usepackage{graphicx}
\usepackage{dcolumn}
\usepackage{bm}
\usepackage{latexsym,epsfig}

\newcommand\ea{\emph{et al. }}

\usepackage{array}
\usepackage{amstext}

\begin{document}

\title{Relativistic general-order coupled-cluster method for high-precision calculations: Application to Al$^+$ atomic clock}
\vspace{0.5cm}

\author{Mih\'aly K\'allay$^{1}$}
\author{B. K. Sahoo$^2$} \author{H. S. Nataraj$^{1}$} \author{B. P. Das$^{3}$} \author{Lucas Visscher$^{4}$}
\affiliation{$^1$ Department of Physical Chemistry and Materials Science, Budapest University of Technology and Economics, Budapest P.O.Box 91, H-1521 Hungary}
\affiliation{$^2$ Theoretical Physics Division, Physical Research Laboratory, Ahmedabad 380009, India}
\affiliation{$^3$ Indian Institute of Astrophysics, 560034 Bangalore, India}
\affiliation{$^4$ Amsterdam Center for Multiscale Modeling,
VU University Amsterdam,
De Boelelaan 1083, 1081 HV Amsterdam, The Netherlands}

\date{\today}

\begin{abstract}
We report the implementation of a general-order relativistic coupled-cluster method for performing 
high-precision calculations of atomic and molecular properties. As a first
application, the static dipole polarizabilities of the ground and first 
excited states of Al$^{+}$ have been determined to precisely estimate the
uncertainty associated with the BBR shift of its clock frequency measurement.
The obtained relative
BBR shift is $-3.66\pm0.44$ for the $3s^2\, ^1S_0^0 \rightarrow 3s3p\, ^3P_0^0$ 
transition in Al$^{+}$ in contrast to the value obtained in the latest 
clock frequency measurement, $-9\pm3$ [Phys. Rev. Lett. {\bf 104}, 070802 (2010)].
The method developed in the present work can be employed to study a variety of subtle effects such as fundamental symmetry 
violations in atoms. 
\end{abstract}

\pacs{31.15.A-,31.15.bw,31.15.V-,32.10.Dk}
\keywords{coupled-cluster method, polarizability, atomic clock}

\maketitle

The role of high precision calculations of various properties of heavy 
atoms and molecules which support the state-of-the-art measurements 
has gained incredible importance in recent years. This is particularly true in the context of atomic clocks 
\cite{rosenband1}, probes of fundamental symmetry 
violations \cite{sahoo,nataraj,porsev}, and search for the variation in the 
fundamental constants \cite{chou}. The relativistic coupled-cluster (CC) method with
single and double excitations (CCSD) supplemented by the important triple
excitations has yielded reasonably accurate results \cite{sahoo,nataraj,porsev}.
However, an extension to this method by including higher order excitations and its application to large systems are extremely challenging. The general nonrelativistic CC 
approach of K\'allay and co-workers provides
one of the most efficient routes to the incorporation of higher order 
excitations by exploiting the features of the many-body diagrammatic techniques
and string algebra \cite{HigherCC}.

In this Letter, we extend the general-order nonrelativistic CC work reported in Ref. \cite{HigherCC} to the
relativistic framework aiming to apply it for high-precision studies in  
several important areas of fundamental physics; to mention a few: atomic clocks,
parity non-conservation (PNC), electric dipole moment (EDM) due to parity and
time reversal violations.
As a proof-of-principle, we have employed the method for the calculation of the 
black-body radiation (BBR) shift in the $3s^2\, ^1S_0^0 \rightarrow 3s3p\, ^3P_0^0$ clock 
transition of Al$^+$. This transition provides the basis for the most accurate 
atomic clock to date \cite{rosenband1,rosenband2,chou}, for which the fractional 
frequency inaccuracy has recently been estimated as
$8.6 \times 10^{-18}$ \cite{chou}. Although the size of the BBR shift in the 
Al$^+$ clock is smaller than those in most of the other ions 
considered for atomic clocks, the associated uncertainty in the 
estimated BBR shift is about 35\% of the total uncertainty. The
BBR shift was also obtained using the static polarizabilities calculated from
the oscillator strengths taken from different sources 
\cite{rosenband}, and it was investigated later by Mitroy {\em et al.} \cite{mitroy}
along similar lines using the configuration interaction (CI) method with a 
semi-empirical core potential. Other nonrelativistic calculations also
provide results that agree with each other \cite{archibong,reshetnikov,hamonou}.

With the general-order relativistic CC method, we also consider the linear
response theory which is the first application of its kind to atoms for the calculation of 
static polarizabilities of the ground and first excited states of Al$^+$.
This method allows for the precise calculations of the ground-state and
one-hole and one-particle excited-state properties, and it can be suitably
modified for applying it to PNC and EDM studies in
the proposed atoms such as Ra and Yb \cite{budker, guest}.

The exact wave function in the CC theory involves an exponential parameterization of the form:
\begin{equation} \label{exp}
| \Psi_{\mathrm {CC}} \rangle = e^{\hat{T}} | 0 \rangle
\end{equation}
where $| 0 \rangle$ is the Dirac-Fock (DF) reference determinant,
and the cluster operator $\hat T$ can be decomposed as
\begin{eqnarray} \label{T}
\hat T \ = \ \sum\limits_{k=1}^n \ \hat T_k
\end{eqnarray}
\begin{eqnarray}
\text{where} \;\;\;\; \hat T_k  = 
\sum\limits_{\stackrel{a_1 < a_2 \dots < a_k}{i_1 < i_2 \dots < i_k}}
t^{a_1 a_2 \dots a_k}_{i_1 i_2 \dots i_k}
 \ a^+_1 i^-_1 a^+_2 i^-_2 \dots a^+_k i^-_k \;\;\;
\end{eqnarray}
The convention followed here is that indices $i$ ($a$) refer to occupied (virtual) 
spinors in the reference determinant.
Projecting onto the excited determinants defined by 
$| \Psi^{a_1 a_2 \dots a_k}_{i_1 i_2 \dots i_k} \rangle =$
$a^+_1 i^-_1 a^+_2 i^-_2 \dots a^+_k i^-_k | 0 \rangle$
we get the nonlinear algebraic equations for the correlation energy $E$ and 
unknown cluster amplitudes for the excitation of any order as
\begin{eqnarray}\label{CCeq}
\langle \Psi^{a_1 a_2 \dots a_k}_{i_1 i_2 \dots i_k} | 
e^{-\hat{T}} \hat{H}_N e^{\hat{T}} | 0 \rangle &=& E \delta_{k,0}, \;\;\;\; (k=1, \dots n) ,
\end{eqnarray}
where $\hat{H}_N$ is the normal-ordered Dirac-Coulomb (DC) Hamiltonian and $k$ 
is the level of excitation. The CC approaches corresponding to the $n = 2, 3, 4, \dots$ values,
i.e., the CC singles and doubles (CCSD), CC singles, doubles, and triples (CCSDT), CC singles, doubles,
triples, and quadruples (CCSDTQ), \dots methods constitute a hierarchy, which converges to the
exact solution of the Dirac equation in the given one-particle basis set.

The excitation energies are obtained invoking the linear-response
CC (LR-CC) theory as given in Refs. \cite{LRT} and \cite{ExcitedCC}.
In LR-CC theory the excitation energies $\omega$ are calculated by determining
the right-hand eigenvalues of the CC Jacobian as
\begin{eqnarray}\label{Right}
\langle \Psi^{a_1 a_2 \dots a_k}_{i_1 i_2 \dots i_k} |
\left[ e^{-\hat{T}} \hat H_N e^{\hat{T}}, \hat R \right] | 0 \rangle \ = \
\omega \ r^{a_1 a_2 \dots a_k}_{i_1 i_2 \dots i_k},
\end{eqnarray}
where operator $\hat R$ has the same structure as the cluster operator with
parameters $r^{a_1 a_2 \dots a_k}_{i_1 i_2 \dots i_k}$.

As it is well-known, the energy of the $i$th state of an atom placed in an isotropic 
electric field of strength $\varepsilon$ changes as 
\begin{eqnarray}\label{polar}
E_i(\varepsilon) = E_i(0) - \frac{\alpha_i}{2} \varepsilon^2 - \dots,
\end{eqnarray}
where $E_i(0)$ and $E_i(\varepsilon)$ are the total energies of the state $i$
in the absence and the presence of the field, respectively, and $\alpha_i$ is the static dipole polarizability of state $i$.
The BBR shift for a transition $|J_i, M_i \rangle \rightarrow |J_j, M_j \rangle$
is the shift of the corresponding transition energy due to the finite background thermal radiation.
At temperature $T$, 
neglecting the dynamic correction factor from the previous finding 
\cite{mitroy}, in the adiabatic expansion it is given by
\begin{eqnarray}
\Delta E_{ij}^\mathrm{BBR} = - \frac{1}{2} (831.9 \ \mathrm{V/m})^2 \left ( \frac{T(\mathrm{K})}{300} \right )^4 (\alpha_i - \alpha_j).
\label{eqn2}
\end{eqnarray}
Consequently the evaluation of the BBR shift
requires the knowledge of the static polarizabilities for the two states
involved in the clock transition. 

It is obvious from Eq. (\ref{polar}) that the
static polarizability can be evaluated as the second derivative of $E_i(\varepsilon)$
with respect to $\varepsilon$. In our study we followed this approach and calculated the polarizabilities
by numerical differentiation. The total energies were 
computed with and without the perturbation; here the perturbation was taken to be $- D \cdot \varepsilon$ where $D$ is the induced
electric dipole moment, and the values of the electric field $\varepsilon$ were fixed to $1 \times 10^{-3}$ and $2\times10^{-3}$ a.u. 
The polarizabilities were obtained from the resulting three energy 
values assuming that they lie on a quartic polynomial. With the test calculations the numerical error of this procedure was found to be negligible.

In oder to approach the exact solution of the Dirac-Coulomb equation for the Al$^+$ ion 
as closely as possible,
the convergent hierarchy of CC methods was combined with the convergent basis
sets in the total energy calculations. The ground-state energies were obtained
using the CCSD, CCSDT, and CCSDTQ methods, while excited-state energies were
determined by the LR-CC method in the same excitation manifold.
The one electron basis sets used were Dunning's triply-augmented correlation
consistent polarized core-valence $X$-tuple-$\zeta$ sets 
\cite{PVXZ2ndRow,PCVXZ2ndRow} abbreviated as t-aug-cc-pCV$X$Z,
where $X$ is the so-called cardinal number of the basis set, $X$ $=$ D, T, Q,
and 5 for double-, triple-, quadruple-, and pentuple-$\zeta$ basis sets,
respectively. The basis sets were uncontracted in all the calculations.
The CC calculations were carried out with our new all-order 
relativistic CC code implemented in the {\sc Mrcc} suite \cite{mrcceng}.
The transformed molecular orbital integrals were generated by the {\sc Dirac} 
package \cite{Dirac}. 

\begin{table}[t]
\caption{\label{Conv} Calculated excitation energies (cm$^{-1}$) and
polarizabilities (a.u.)}
\begin{tabular}{lcccc} \hline\hline
&Excitation &\multicolumn{3}{c}{Polarizability} \\
\cline{3-5}
&energy     & Ground state & Excited state & Differential \\ \hline
\multicolumn{5}{l}{t-aug-cc-pCVDZ (1s, 2s, and virtuals $>$ 5E$_h$ are frozen)}\\
CCSD  & 37222& 24.215& 24.380& 0.165 \\
CCSDT & 37324& 24.158& 24.357& 0.199 \\
CCSDTQ& 37326& 24.156& 24.358& 0.202 \\
      &      &       &       &       \\
\multicolumn{5}{l}{t-aug-cc-pCVDZ}   \\
CCSD  & 37005& 24.203& 24.261& 0.058 \\
CCSDT & 37167& 24.072& 24.208& 0.136 \\
      &      &       &       &       \\
\multicolumn{5}{l}{t-aug-cc-pCVTZ}   \\
CCSD  & 37228& 24.143& 25.040& 0.897 \\
CCSDT & 37373& 24.017& 24.979& 0.962 \\
      &      &       &       &       \\
\multicolumn{5}{l}{t-aug-cc-pCVQZ}   \\
CCSD  & 37160& 24.273& 24.700& 0.427 \\
      &      &       &       &       \\
\multicolumn{5}{l}{t-aug-cc-pCV5Z}   \\
CCSD  & 37186& 24.251& 24.656& 0.406 \\ \hline\hline
\end{tabular}
\end{table}

\begin{table*}[ht]
\caption{\label{Comp} Composite excitation energies (cm$^{-1}$), 
polarizabilities (a.u.), and their estimated errors. Note that 1s, 2s, and the virtual orbitals above 5 E$_h$ were frozen for the CCSDTQ-CCSDT calculations.}
\begin{tabular}{lrrrrc} \hline\hline
Contribution &\multicolumn{1}{c}{Excitation}&\multicolumn{3}{c}{Polarizability}& Source \\
\cline{3-5}
             &\multicolumn{1}{c}{energy}    & Ground state    & Excited state  &\multicolumn{1}{c}{Differential}         &        \\ \hline
CCSD         &37186$\pm$25& 24.251$\pm$0.022& 24.656$\pm$0.044& 0.406$\pm$0.021& t-aug-cc-pCV5Z \\
CCSDT-CCSD   &  146$\pm$33& $-$0.126$\pm$0.011& $-$0.061$\pm$0.015& 0.065$\pm$0.026& t-aug-cc-pCVTZ \\
CCSDTQ-CCSDT &    2$\pm$4 & $-$0.002$\pm$0.005&  0.001$\pm$0.002& 0.003$\pm$0.007& t-aug-cc-pCVDZ \\ 
Breit+QED    &  $-$6$\pm$6 & 0.015$\pm$0.015& 0.018$\pm$0.018 & 0.003$\pm$0.003& Numerical MCDF \\ \hline         
Composite    &37326$\pm$68& 24.137$\pm$0.053& 24.614$\pm$0.078& 0.477$\pm$0.057& Sum of all contributions\\ \hline\hline
\end{tabular}
\end{table*}

To give an accurate estimate of the properties we studied,
we adopted a composite scheme, which is
well-established in quantum chemistry and widely used for 
highly-accurate calculations of molecular properties (see, e.g., Refs.
\cite{W3W4,HEAT,EqGeomII,W4,Feller2008}), in which the calculations with a
particular method in the CC hierarchy are carried out with the largest
possible basis set and the largest possible number of correlated
electrons. In practice, CCSD and CCSDT
calculations were performed with pentuple- and triple-$\zeta$
basis sets, respectively, correlating all electrons and all orbitals. 
CCSDTQ calculations were only feasible with the t-aug-cc-pCVDZ basis 
set, but further approximations were necessary even in this basis, and
the 1s and 2s electrons were frozen as well as the virtual orbitals
lying above 5 E$_h$ were dropped. Our final estimates were obtained by
adjusting the pentuple-$\zeta$ CCSD values with the CCSDT-CCSD and
CCSDTQ-CCSDT increments computed with the triple- and double-$\zeta$
basis sets, respectively. The error of our computed values were 
estimated on the basis of the convergence pattern of the results. To improve
the results further, we
estimated the contributions from Breit interaction and QED corrections
using the numerical multi-configurational DF (MCDF)
method as implemented in the {\sc Mcdfgme} program \cite{MCDFGME} and
the sum-over-states expression for polarizabilities \cite{mitroy}.

The calculated polarizabilities are compiled in Table \ref{Conv} where 
we also present the excitation energy of the clock transition. Since
the latter is precisely known from experiments, the performance of our 
approach can be partly judged from the agreement of our calculated and
the measured excitation energy.

The convergence of both the polarizabilities and excitation energies
with increasing levels of correlation is rapid. The CCSD values themselves are 
reliable; further, the contribution of triple excitations to both 
properties is less than 1\%. Interestingly the polarizability of the 
ground-state is more sensitive to correlation effects than the 
excited-state: the effect of triple excitations for the ground state is 
twice as large as that for the excited state, viz. 0.13 a.u. vs. 0.06 
a.u. The
magnitude of the triples contribution to the polarizability shift is 
also moderate, it only amounts to 0.06 a.u., however, it is more than
10\% of the composite value and thus cannot be ignored. The effect of
quadruple excitations is approximately two orders of magnitude smaller 
than that
of the triples and can be considered as negligible, which also implies
that higher-order correlation contributions can safely be ignored.

The basis set convergence of the properties we have studied is in accordance
with the usual trend---relatively slow, but the results are close to 
the basis set limit
when large basis sets are employed. The polarizabilities and 
excitation energies are already reliable in the smaller basis sets,
while the polarizability shift, which is a small difference of two
large numbers, requires at least quadruple-$\zeta$-quality basis set
even for a qualitatively correct result. It is interesting to note that
in this case the polarizability of the excited-state is more sensitive 
to the quality of the basis set than the ground 
state. From the comparison of the
quadruple- and pentuple-$\zeta$ results we observe that the CCSD
excitation energies and polarizabilities change on the scales of 10 cm$^{-1}$ 
and 0.01 a.u., respectively, which means that the relative change is 
about 0.1\% for both properties. Since the basis-set error 
decreases monotonically with the size of the basis set, the error with
respect to the infinite basis set limit is also expected to be less
than 0.1\%. Unfortunately the errors of the ground- and excited-state
polarizabilities do not cancel each other, and consequently the absolute
error of the CCSD polarizability shift is larger. For the aforementioned
reason its relative error is also significantly larger, a couple of 
percent of the total value.
Similar conclusions can be drawn for the contribution of triple 
excitations. The CCSDT-CCSD difference also changes in the 10 cm$^{-1}$ 
and 0.01 a.u. range for excitation energies and polarizabilities,
respectively, when going from the double- to the triple-$\zeta$ basis
set, and the change in the polarizability shift is only 0.013 a.u.
Thus the error in our final estimates stemming from the calculations
of the triples contribution is also smaller than 0.1\% (3\%) for the
excitation energy and polarizabilities (polarizability shift).
\begin{table}[ht]
\caption{\label{Compare} Comparison of theoretical and experimental
polarizabilities (a.u.), and relative BBR shifts.}
\begin{tabular}{lllc}\hline\hline
\multicolumn{2}{c}{Polarizability}&\multicolumn{1}{c}{BBR shift} & Reference \\
\cline{1-2}
   $3s^2\, ^1S_0^0$ & $3s3p\, ^3P_0^0$   &\multicolumn{1}{c}{$\times10^{18}$} & \\ \hline
   24.19          &                &                  & \cite{archibong} \\
24.83$\pm$5.26 & 24.63$\pm$4.93  &  $-$8$\pm$3  & \cite{rosenband} \\
 24.20$\pm$0.75 &                    &       & \cite{reshetnikov}         \\
   24.12          &                     &    & \cite{StatPolMg}            \\
    24.14$\pm$0.12 & 24.62$\pm$0.25 & $-$4.18$\pm$3.18 & \cite{mitroy}\\
   24.22$\pm$1.21 & 24.78$\pm$1.24 & $-$4.3$\pm$2.5 & \cite{Rollin} \\
24.14$\pm$0.05 & 24.61$\pm$0.08 & $-$3.66$\pm$0.44 & This work \\ \hline\hline
\end{tabular}
\end{table}

The calculation of the properties that have been investigated using the composite
approach outlined above is shown in Table \ref{Comp}
in detail, where we also present our error estimates based on the
convergence of the contributions with the basis set. It has been found
in numerous studies that in the higher members of correlation-consistent
basis set
family the basis set error for various properties is usually reduced by 
a factor of at least two when increasing the cardinal number of the 
basis set by one. The reduction of the basis set error would also be 
valid for the current properties. In fact, the ratio of the 
quadruple-$\zeta$$-$triple-$\zeta$ and 
pentuple-$\zeta$$-$quadruple-$\zeta$ differences of CCSD excitation 
energies, ground- and excited-state polarizabilities, and polarizability
shifts is 2.7, 5.8, 7.8, and 22.1, respectively. Thus we presume that 
the entire difference between the
pentuple- and quadruple-$\zeta$ results is a conservative estimate for
the basis set error of the CCSD values, and we attach these numbers as
error bars. The estimation of the intrinsic error of the CCSDT-CCSD
contributions is less straightforward since the results are not 
available in the larger basis sets. Therefore we take twice the difference
between the double- and the triple-$\zeta$ triples contributions
as a conservative choice. The quadruples contribution, i.e.,
the CCSDTQ-CCSDT difference is only available in one basis set, and no
conclusion about its basis set dependence can be drawn. Consequently
we take twice the entire contribution as the error bar. The contribution of Breit and QED corrections for the excitation energy is $\sim -6$ cm$^{-1}$ while for polarizabilities of the ground and excited states it is 0.015 and 0.018 a.u., respectively. We would like to remark that these effects are computed using numerical orbitals at the DF level of the theory and hence they are devoid of any basis set incompleteness errors. As the missing correlation contribution to these effects is not expected to exceed its DF value, we have taken the entire value itself as the upper limit of the error.

For the excitation energy a highly-accurate experimental value, 37393$\pm$0 cm$^{-1}$ is 
available \cite{rosenband1}, thus the agreement between the experimental and our best calculated
excitation energy, 37326$\pm$68 cm$^{-1}$ is very good and the deviation is within 0.2\% of the experimental energy. 

We compare our polarizabilities and the BBR shift to the previous 
theoretical and empirical results in Table \ref{Compare}. Our results are in
good agreement with the previous computational results, however, more accurate 
than the latter. In contrast, there is a considerable discrepancy between 
the present and the experimental BBR shift. There is a brief discussion on 
various approaches employed to calculate the polarizabilities and the BBR shift
by Mitroy \ea \cite{mitroy}, hence we do not repeat them here, however we would like to emphasize that
our results are the first {\it ab initio} values based on a relativistic framework.

In conclusion, we have developed a general-order relativistic coupled-cluster method for 
high-precision calculations in atoms and molecules.
Using this method the ground-state, excited-state, and differential
polarizabilities of the Al$^+$ ion are obtained to be 24.14$\pm$0.05, 24.61$\pm$0.08, and 0.48$\pm$0.06 a.u., 
respectively. From the latter value and the measured clock 
frequency of $1.121015393207851\times10^{-15}$ Hz \cite{chou}
we obtain $-$0.0041$\pm$0.0005 Hz for the absolute and $-$3.66$\pm$0.44 for 
the relative BBR shift. It is the most accurate estimate of the BBR shift in
Al$^+$, using which the 
systematic shift in the above frequency measurement can be obtained as 
$(-1112.46\pm6.04)\times10^{-18}$ against the value of
$(-1117.8\pm8.6)\times10^{-18}$ considered in Ref. \cite{chou}.

Financial support to M.K. has
been provided by the European Research Council (ERC) under FP7,
ERC Grant Agreement No. 200639, and by the Hungarian Scientific
Research Fund (OTKA), Grant No. NF72194. M.K. and B.P.D. acknowledge
the Indo-Hungarian (IND 04/2006) project. M.K. acknowledges the
Bolyai Research Scholarship of the Hungarian Academy of Sciences.
B.K.S. thanks T. Rosenband for useful discussions.
L.V. has been supported by NWO through the VICI programme.


\end{document}